\documentstyle[twoside]{article}
\input{epsf.sty}

\catcode`\@=11
\long\def\@makefntext#1{
\protect\noindent \hbox to 3.2pt {\hskip-.9pt  
$^{{\eightrm\@thefnmark}}$\hfil}#1\hfill}		

\def\@makefnmark{\hbox to 0pt{$^{\@thefnmark}$\hss}}	
	
\def\ps@myheadings{\let\@mkboth\@gobbletwo
\def\@oddhead{\hbox{}
\rightmark\hfil\eightrm\thepage}   
\def\@oddfoot{}\def\@evenhead{\eightrm\thepage\hfil
\leftmark\hbox{}}\def\@evenfoot{}
\def\sectionmark##1{}\def\subsectionmark##1{}}

\oddsidemargin=\evensidemargin
\newcounter{sectionc}\newcounter{subsectionc}\newcounter{subsubsectionc}
\renewcommand{\section}[1] {\vspace{12pt}\addtocounter{sectionc}{1} 
\setcounter{subsectionc}{0}\setcounter{subsubsectionc}{0}\noindent 
	{\tenbf\thesectionc. #1}\par\vspace{5pt}}
\renewcommand{\subsection}[1] {\vspace{12pt}\addtocounter{subsectionc}{1} 
	\setcounter{subsubsectionc}{0}\noindent 
	{\bf\thesectionc.\thesubsectionc. {\kern1pt \bfit #1}}\par\vspace{5pt}}
\renewcommand{\subsubsection}[1] {\vspace{12pt}\addtocounter{subsubsectionc}{1}
	\noindent{\tenrm\thesectionc.\thesubsectionc.\thesubsubsectionc.
	{\kern1pt \tenit #1}}\par\vspace{5pt}}
\newcommand{\nonumsection}[1] {\vspace{12pt}\noindent{\tenbf #1}
	\par\vspace{5pt}}

\newcounter{appendixc}
\newcounter{subappendixc}[appendixc]
\newcounter{subsubappendixc}[subappendixc]
\renewcommand{\thesubappendixc}{\Alph{appendixc}.\arabic{subappendixc}}
\renewcommand{\thesubsubappendixc}
	{\Alph{appendixc}.\arabic{subappendixc}.\arabic{subsubappendixc}}

\renewcommand{\appendix}[1] {\vspace{12pt}
        \refstepcounter{appendixc}
        \setcounter{figure}{0}
        \setcounter{table}{0}
        \setcounter{lemma}{0}
        \setcounter{theorem}{0}
        \setcounter{corollary}{0}
        \setcounter{definition}{0}
        \setcounter{equation}{0}
        \renewcommand{\thefigure}{\Alph{appendixc}.\arabic{figure}}
        \renewcommand{\thetable}{\Alph{appendixc}.\arabic{table}}
        \renewcommand{\theappendixc}{\Alph{appendixc}}
        \renewcommand{\thelemma}{\Alph{appendixc}.\arabic{lemma}}
        \renewcommand{\thetheorem}{\Alph{appendixc}.\arabic{theorem}}
        \renewcommand{\thedefinition}{\Alph{appendixc}.\arabic{definition}}
        \renewcommand{\thecorollary}{\Alph{appendixc}.\arabic{corollary}}
        \renewcommand{\theequation}{\Alph{appendixc}.\arabic{equation}}
        \noindent{\tenbf Appendix \theappendixc #1}\par\vspace{5pt}}
\newcommand{\subappendix}[1] {\vspace{12pt}
        \refstepcounter{subappendixc}
        \noindent{\bf Appendix \thesubappendixc. {\kern1pt \bfit #1}}
	\par\vspace{5pt}}
\newcommand{\subsubappendix}[1] {\vspace{12pt}
        \refstepcounter{subsubappendixc}
        \noindent{\rm Appendix \thesubsubappendixc. {\kern1pt \tenit #1}}
	\par\vspace{5pt}}

\newcommand{\textlineskip}{\baselineskip=13pt}
\newcommand{\smalllineskip}{\baselineskip=10pt}

\def\eightcirc{
\begin{picture}(0,0)
\put(4.4,1.8){\circle{6.5}}
\end{picture}}
\def\eightcopyright{\eightcirc\kern2.7pt\hbox{\eightrm c}} 

\newcommand{\copyrightheading}[1]
	{\vspace*{-2.5cm}\smalllineskip{\flushleft
	{\footnotesize International Journal of Modern Physics C, #1}\\
	{\footnotesize $\eightcopyright$\, World Scientific Publishing
	 Company}\\
	 }}

\newcommand{\publisher}[2]{{\begin{center}\footnotesize\smalllineskip 
	Received #1\\
	Revised #2
	\end{center}
	}}

\def\abstracts#1#2#3{{
	\centering{\begin{minipage}{4.5in}\baselineskip=10pt\footnotesize
	\parindent=0pt #1\par 
	\parindent=15pt #2\par
	\parindent=15pt #3
	\end{minipage}}\par}} 

\renewenvironment{thebibliography}[1]
        {\frenchspacing
	 \ninerm\baselineskip=11pt
         \begin{list}{\arabic{enumi}.}
        {\usecounter{enumi}\setlength{\parsep}{0pt}     
	 \setlength{\leftmargin 12.7pt}{\rightmargin 0pt} 
         \setlength{\itemsep}{0pt} \settowidth
	{\labelwidth}{#1.}\sloppy}}{\end{list}}

\newcounter{itemlistc}
\newcounter{romanlistc}
\newcounter{alphlistc}
\newcounter{arabiclistc}

\newcommand{\fcaption}[1]{
        \refstepcounter{figure}
        \setbox\@tempboxa = \hbox{\footnotesize Fig.~\thefigure. #1}
        \ifdim \wd\@tempboxa > 5in
           {\begin{center}
        \parbox{5in}{\footnotesize\smalllineskip Fig.~\thefigure. #1}
            \end{center}}
        \else
             {\begin{center}
             {\footnotesize Fig.~\thefigure. #1}
              \end{center}}
        \fi}

\newcommand{\tcaption}[1]{
        \refstepcounter{table}
        \setbox\@tempboxa = \hbox{\footnotesize Table~\thetable. #1}
        \ifdim \wd\@tempboxa > 5in
           {\begin{center}
        \parbox{5in}{\footnotesize\smalllineskip Table~\thetable. #1}
            \end{center}}
        \else
             {\begin{center}
             {\footnotesize Table~\thetable. #1}
              \end{center}}
        \fi}

\def\@citex[#1]#2{\if@filesw\immediate\write\@auxout
	{\string\citation{#2}}\fi
\def\@citea{}\@cite{\@for\@citeb:=#2\do
	{\@citea\def\@citea{,}\@ifundefined
	{b@\@citeb}{{\bf ?}\@warning
	{Citation `\@citeb' on page \thepage \space undefined}}
	{\csname b@\@citeb\endcsname}}}{#1}}

\newif\if@cghi
\def\cite{\@cghitrue\@ifnextchar [{\@tempswatrue
	\@citex}{\@tempswafalse\@citex[]}}
\def\citelow{\@cghifalse\@ifnextchar [{\@tempswatrue
	\@citex}{\@tempswafalse\@citex[]}}
\def\@cite#1#2{{$\null^{#1}$\if@tempswa\typeout
	{IJCGA warning: optional citation argument 
	ignored: `#2'} \fi}}

\def\pmb#1{\setbox0=\hbox{#1}
	\kern-.025em\copy0\kern-\wd0
	\kern.05em\copy0\kern-\wd0
	\kern-.025em\raise.0433em\box0}

\def\fnt#1#2{\footnotetext{\kern-.3em
	{$^{\mbox{\scriptsize #1}}$}{#2}}}

\def\fpage#1{\begingroup
\voffset=.3in
\thispagestyle{empty}\begin{table}[b]\centerline{\footnotesize #1}
	\end{table}\endgroup}

\def\runninghead#1#2{\pagestyle{myheadings}
\markboth{{\protect\footnotesize\it{\quad #1}}\hfill}
{\hfill{\protect\footnotesize\it{#2\quad}}}}
\font\tenrm=cmr10
\font\tenit=cmti10 
\font\tenbf=cmbx10
\font\bfit=cmbxti10 at 10pt
\font\ninerm=cmr9

\font\eightrm=cmr8

\def\qed{\hbox{${\vcenter{\vbox{			
   \hrule height 0.4pt\hbox{\vrule width 0.4pt height 6pt
   \kern5pt\vrule width 0.4pt}\hrule height 0.4pt}}}$}}

\def\bsc{{\sc a\kern-6.4pt\sc a\kern-6.4pt\sc a}}  	
\def\bflatex{\bf L\kern-.30em\raise.3ex\hbox{\bsc}\kern-.14em 
T\kern-.1667em\lower.7ex\hbox{E}\kern-.125em X} 

\begin{document}

\runninghead{Improved spin dynamics simulations...} {Improved spin dynamics
simulations...}

\normalsize\textlineskip
\thispagestyle{empty}
\setcounter{page}{1}

\copyrightheading{}			

\vspace*{0.88truein}

\fpage{1}

\centerline{\bf IMPROVED SPIN DYNAMICS SIMULATIONS}
\vspace*{0.035truein}
\centerline{\bf  OF MAGNETIC EXCITATIONS}
\vspace*{0.37truein}
\centerline{\footnotesize D. P. LANDAU, SHAN-HO TSAI, M. KRECH,
and ALEX BUNKER\footnote{
permanent address:  Max Planck Institut fur Polymerforschung, Mainz, Germany}}
\vspace*{0.015truein}
\centerline{\footnotesize\it Center for Simulational Physics, The University
of Georgia,}
\baselineskip=10pt
\centerline{\footnotesize\it Athens, GA 30602, U.S.A.}

\vspace*{0.21truein}
\publisher{(received date)}


\vspace*{0.21truein}
\abstracts{
Using Suzuki-Trotter decompositions of exponential operators we
describe new algorithms for the numerical integration of the equations
of motion for classical spin systems. These techniques conserve spin
length exactly and, in special cases, also conserve the energy and
maintain time reversibility. We investigate integration schemes of up to
eighth order and show that these new algorithms can be used with
much larger time steps than a well established predictor-corrector method.
These methods may lead to a substantial speedup of spin dynamics
simulations, however, the choice of
which order method to use is not always straightforward.
}{}{}

\vspace*{1pt}\textlineskip
\section{Introduction}
\vspace*{-0.5pt}
\noindent
Our understanding of static behavior near phase transitions is now
mature and has resulted largely from the
investigation of simple model spin systems such as the Ising,
the XY, and the Heisenberg model. These models are equally
valuable for the investigation
of dynamic critical behavior and dynamic scaling. 
Realistic models of magnetic materials can be constructed from these
simple spin models, however, the theoretical analysis of
experimentally accessible quantities, such as the dynamic structure
factor, is usually too demanding for analytical methods. Computer
simulations are
beginning to provide important information about dynamic critical
behavior and material properties of model magnetic systems
\cite{EvLan96,ChenLan94,BunChenLan96}. These simulations use
model Hamiltonians with continuous degrees of freedom represented by a
three-component spin ${\bf S}_k$ with fixed length $|{\bf S}_k| = 1$
for each lattice site $k$. A typical model Hamiltonian is then given
by
\begin{equation}
\label{H}
{\cal H} = -J \sum_{<k,l>} \left(S_k^x S_l^x + S_k^y S_l^y + \lambda
S_k^z S_l^z \right) - D \sum_k \left( S_k^z \right)^2 ,
\end{equation}
where $J$ is the exchange integral, $<k,l>$ denotes a nearest-neighbor
pair of spins ${\bf S}_k$, $\lambda$ is an anisotropy parameter, and $D$
determines the strength of a single-site or crystal field anisotropy.

Modelling specific magnetic materials may require
additional interactions in the Hamiltonian, such as
two-spin exchange interactions between more distant neighbors \cite{Bohm80}, 
three spin exchange interactions, or even biquadratic coupling \cite{BEG71}.

The thermodynamic properties can be obtained from a Monte-Carlo simulation 
and the {\em dynamical} properties of the spin system are provided by 
solutions to the equations of motion given by \cite{EvLan96,ChenLan94,BunChenLan96}

\begin{equation}
\label{eqmot}
{d \over dt} {\bf S}_k = {\partial {\cal H} \over \partial {\bf
S}_k} \times {\bf S}_k
\end{equation}

\noindent
These equations must be integrated numerically, where a Monte-Carlo simulation of the
model provides {\em equilibrium} configurations as initial conditions
for Eq.(\ref{eqmot}). The most important quantity to be extracted from the
numerical results is the dynamic structure factor $S({\bf
q},\omega)$, which is given by the space-time Fourier transform
of the spin-spin correlation function
\begin{equation}
\label{correl}
{\cal G}^{\alpha,\beta}({\bf r}_k - {\bf r}_l, t - t') \equiv \langle
S_k^{\alpha} (t) S_l^{\beta}(t') \rangle ,
\end{equation}
where $\alpha,\beta = x,y,z$, ${\bf r}_k$
and ${\bf r}_l$ are lattice vectors, and the average $\langle \dots
\rangle$ must be taken over a large number of independent
initial {\em equilibrium} configurations.  (This procedure is
appropriate since the typical
time over which Eq.(\ref{eqmot}) can be integrated is much shorter than
typical timescales set by other excitations.)

To speed up the numerical integration of Eq.(\ref{eqmot}) it is desirable 
to use the largest possible time step; however, with
standard methods the size of the time step is severely limited
by the accuracy within which the {\em conservation laws} of the dynamics
are obeyed. It is evident from Eq.(\ref{eqmot}) that the
total energy is conserved, and if, for example, $D = 0$ and $\lambda = 1$ (isotropic
Heisenberg model) the magnetization ${\bf M} = \sum_k {\bf S}_k$ is also
conserved. For the anisotropic Heisenberg model, i.e.,
$\lambda \neq 1$ or $D \neq 0$ only $M_z$
is conserved. Conservation of spin length and energy
is particularly crucial, because the condition $|{\bf S}_k| = 1$ is a
major part of the definition of the model and the energy of a
configuration determines its statistical weight. It would therefore
also be desirable to devise an algorithm which conserves these two
quantities {\em exactly}.

In the remaining sections we describe a 4th-order predictor-corrector 
method and a new integration procedure, which is based on Trotter-Suzuki 
decompositions of exponential operators \cite{FraHuaLei,KreBunLan}. 
We compare both schemes with special regard
to speed and the accuracy within which the conservation laws hold.

\section{Integration Methods}
\subsection{Predictor-corrector methods}
Predictor-corrector methods have been quite effective for the
numerical integration of spin equations of motion; however, in order
to limit truncation errors small time steps $\delta t$ must be used
with at least a fourth-order scheme.  In a more symbolic form 
Eq.(\ref{eqmot}) can be written as $\dot{y} =
f(y)$ with the initial condition $y(0) = y_0$, where $y$ is a
short-hand notation of a complete spin configuration.  The initial
equilibrium configuration is denoted by $y_0$. The predictor
step of the scheme used here, the explicit Adams-Bashforth
four-step method \cite{NumAna81}, is
\begin{equation}
\label{AdBash}
y(t+\delta t) = y(t) + {\delta t \over 24} \left[55 f(y(t)) - 59
f(y(t-\delta t)) + 37 f(y(t-2\delta t)) - 9 f(y(t-3\delta t)) \right]
\end{equation}
which has a local truncation error of the order $(\delta t)^5$. The
corrector step consists of typically one iteration of the implicit
Adams-Moulton three-step method \cite{NumAna81}
\begin{equation}
\label{AdMoul}
y(t+\delta t) = y(t) + {\delta t \over 24} \left[9 f(y(t+\delta t)) +
19 f(y(t)) - 5 f(y(t-\delta t)) + f(y(t-2\delta t)) \right]
\end{equation}
which also has a local truncation error of the order $(\delta t)^5$.
(Values for $y(\delta t)$, $y(2\delta t)$, and $y(3\delta t)$ in addition
to $y(0) = y_0$ can be provided by three successive integrations of 
$\dot{y} = f(y)$ (see Eq.(\ref{eqmot})) by the 4th-order Runge-Kutta 
method \cite{NumAna81} .)  This method requires that spin
configurations at the last four time steps must be kept in memory.

This predictor-corrector method is very general and is independent of
the special structure of the right-hand side of the equations of
motion (see Eq.(\ref{eqmot})). The conservation laws discussed earlier
will only be observed within the accuracy set by the
truncation error of the method. In practice, this limits the time step
to typically $\delta t = 0.01/J$ in $d = 3$ for
the isotropic model $(D = 0)$ \cite{BunChenLan96}, where
the total integration time is typically $600/J$ or less.

\subsection{Suzuki-Trotter decomposition methods}
The motion of a spin may be viewed
as a precession of the spin ${\bf S}$ around an effective axis
$\Omega$ which is itself time dependent.  The lattice
can be decomposed into two sublattices such that a spin on one
sublattice precesses in a local field $\Omega$ of neighbor
spins which are {\em all} located on the other sublattice. For the
Hamiltonian in Eq.(\ref{H}) there are only two such sublattices
if the underlying lattice is simple cubic.

To illustrate the method, we consider first the $D=0$ case.
The basic idea of the algorithm is to rotate a spin about its local field
$\Omega$ by an angle $\alpha = |\Omega|\delta t$, rather than directly
integrate Eq.(\ref{eqmot}). This procedure
guarantees the conservation of the spin length $|{\bf S}|$
and energy to within machine accuracy.
 Denoting the two sublattices by $\cal A$ and $\cal B$,
respectively, we can express the local fields acting on the
spins on sublattice $\cal A$ and $\cal B$ as 
$\Omega_{\cal A}[\{{\bf S}\}]$ and $\Omega_{\cal B}[\{{\bf S}\}]$, 
respectively. The set of equations of motion for spins on one sublattice 
reduces to a {\em linear} system of
differential equations if the spins on the other sublattice are kept
fixed. Thus an {\em alternating} update scheme may be used, i.e., the
spins ${\bf S}_{k \in {\cal A}}$ are rotated for the given values of
${\bf S}_{k \in {\cal B}}$ and vice versa. (The
scalar products ${\bf S}_{k \in {\cal A}} \cdot \Omega_{\cal A}[\{{\bf
S}\}]$ remain constant during the update of ${\bf S}_{k \in {\cal A}}$
and the scalar products ${\bf S}_{k \in {\cal B}} \cdot \Omega_{\cal
B}[\{{\bf S}\}]$ remain constant during the update of ${\bf S}_{k \in
{\cal B}}$).  Note, that each
sublattice rotation is performed with the {\em actual} values of the
spins on the other sublattice, so that only a {\em single} copy of the
spin configuration is kept in memory at any time. However, the
magnetization will not be conserved during the above rotation
operations. Since the two alternating rotation operations do not
commute, a closer examination of the sublattice decomposition
of the spin rotation is required.

We now decompose a full configuration into two sublattice components $y_{\cal A}$
and $y_{\cal B}$, i.e. $y = (y_{\cal A},y_{\cal B})$, and denote 
by matrices $A$ and $B$ the
generators of the rotation of the spin configuration $y_{\cal A}$ on
sublattice $\cal A$ at fixed $y_{\cal B}$ and of the spin
configuration $y_{\cal B}$ on sublattice $\cal B$ at fixed $y_{\cal
A}$, respectively. The update of the configuration $y$ from time $t$
to $t + \delta t$ is then given by an exponential (matrix)
operator
\begin{equation}
\label{eAB}
y(t+\delta t) = e^{(A + B)\delta t} y(t) .
\end{equation}
Although the exponential operator in Eq.(\ref{eAB}) rotates each spin
of the configuration it has no simple explicit form, because the
rotation axis for each spin depends on the configuration itself; however, the
operators $e^{A \delta t}$ and $e^{B \delta t}$ which rotate $y_{\cal
A}$ at fixed $y_{\cal B}$ and $y_{\cal B}$ at fixed $y_{\cal A}$,
respectively, {\em do} have a simple explicit form. We demonstrate
this for the case $\lambda = 1$ and $D = 0$ in Eq.(\ref{H}). For each
$k \in {\cal A}$ we find
\begin{equation}
\label{OmegaA}
\Omega_{\cal A}[\{{\bf S}\}] = -J \sum_{l = NN(k)} {\bf S}_l \equiv
\Omega_k ,
\end{equation}
where $NN(k)$ denotes the nearest neighbors of $k$ (which belong to
$y_{\cal B}$). Eq.(\ref{OmegaA}) can be readily generalized for
$\lambda \neq 1$, the case $D \neq 0$ will be discussed below. The
explicit rotation of spins on each sublattice reads
(see also Watson {\it et al} \cite{WatBluVin69})
\begin{equation}
\label{Sktdt}
{\bf S}_k(t+\delta t) = {\Omega_k (\Omega_k \cdot {\bf S}_k(t)) \over
\Omega_k^2} + \left[ {\bf S}_k(t) - {\Omega_k (\Omega_k \cdot {\bf
S}_k(t)) \over \Omega_k^2} \right] \cos (|\Omega_k|\delta t) +
{\Omega_k \times {\bf S}_k(t) \over |\Omega_k|} \sin (|\Omega_k|\delta t).
\end{equation}
Note that according to Eq.(\ref{Sktdt}) $\Omega_k \cdot {\bf S}_k(t +
\delta t) = \Omega_k \cdot {\bf S}_k(t)$ and
energy is thus conserved. The alternating update scheme for the
integration of the equations of motion
amounts to the replacement $e^{(A + B) \delta t} \to e^{A \delta t}
e^{B \delta t}$ in Eq.(\ref{eAB}), which is only correct \cite{SuzUme93}
up to order $(\delta t)^2$. The magnetization will
therefore only be conserved up to terms of the order $\delta t$ (global
truncation error), but one can employ higher order
Suzuki-Trotter decompositions of the exponential operator in
Eq.(\ref{eAB}) to decrease the local truncation error of the algorithm
and thus improve the conservation.  The simplest possible
improvement is given by the 2nd-order decomposition\cite{SuzUme93}
\begin{equation}
\label{eA2BA2}
e^{(A + B) \delta t} = e^{A \delta t / 2} e^{B \delta t} e^{A \delta t
/ 2} + {\cal O}(\delta t^3).
\end{equation}
which is equivalent to the midpoint integration
method applied to the equations of motion (see also Watson {\it et al}
\cite{WatBluVin69}). 
We can also use the mth-order decomposition \cite{SuzUme93}
\begin{equation}
\label{epABA}
e^{(A + B) \delta t} = \prod_{i=1}^n e^{p_i A \delta t / 2} e^{p_i B
\delta t} e^{p_i A \delta t / 2} + {\cal O}(\delta t^{m+1})
\end{equation}
where $n=5$ for 4th-order and $n=15$ for 8th-order and the parameters
$p_i$ are given by Suzuki and Umeno \cite{SuzUme93}. 

The additional computational effort needed to evaluate higher order
expressions can be compensated to some extent
by using larger time steps. The evaluation of the
trigonometric functions in Eq.(\ref{Sktdt}) can also be avoided 
since the above decompositions are only correct to within a
certain order in $\delta t$ and it is therefore sufficient to replace
$\sin x$ and $\cos x$ by appropriate Taylor polynomials; alternatively a 
Cayley transform could be used (up to 2nd and 4th-order, it corresponds
to $\sin x=x\; p(x)/[p^2(x)+(x/2)^2]$, with $p(x)=1$ and $p(x)=1-x^2/12$,
respectively; determining $\cos x$ from $\sin^2 x + \cos^2 x=1$
ensures spin-length conservation). For a 4th-order method the Cayley transform
was $10 - 20 \%$ faster, depending upon the machine.
Note that the decompositions maintain the
time inversion property of $e^{(A + B) \delta t}$.  The inclusion of
next-nearest neighbor bilinear interactions on a simple cubic lattice
can be treated within the above
framework if the lattice is decomposed into four sublattices.

This approach can also be extended to the case $D \neq 0$, but in
contrast to the isotropic case, the equation of
motion for each individual spin on each sublattice is {\em nonlinear}.
In practice, the best form of solution is via iterative numerical methods.  
For the sublattice decomposition of the spin rotation the requirement for 
energy conservation in the presence of a single site anisotropy is
\begin{equation}
\label{energyD}
\Omega_k \cdot {\bf S}_k(t+\delta t) - D \left[S_k^z(t+\delta t)\right]^2
 = \Omega_k \cdot {\bf S}_k(t) - D \left[S_k^z(t)\right]^2
\end{equation}
for $k \in {\cal A}$ and $k \in {\cal B}$, where $\Omega_k$ is given
by Eq.(\ref{OmegaA}). In order to perform a rotation operation in
analogy to Eq.(\ref{Sktdt}) we have to identify an effective rotation
axis. This can be achieved by rewriting Eq.(\ref{energyD}) in the form
$\widetilde{\Omega}_k \cdot ({\bf S}_k(t + \delta t) - {\bf S}_k(t)) =
0$, where
\begin{equation}
\label{OmegaD}
\widetilde{\Omega}_k = \Omega_k - D \left(0,0,S_k^z(t) +
S_k^z(t+\delta t) \right).
\end{equation}
Since the rotation requires knowledge of $S_k^z$ at the future time $t +
\delta t$, this problem can be
solved iteratively starting from the initial value $S_k^z(t+\delta t)
= S_k^z(t) + (\Omega_k\times S_k(t))^z \delta t$ in Eq.(\ref{OmegaD}) 
and performing several updates according
to the decompositions given by Eqs.(\ref{eA2BA2}) or (\ref{epABA}),
respectively, in order to improve energy conservation according to
Eq.(\ref{energyD}). 
Both the degree of conservation and the execution time depend to some
extent on the number of iterations used.
The initial value for $S_k^z(t+\delta t)$ used here yields a better energy
conservation (at almost no extra CPU time) than using the initial value
$S_k^z(t+\delta t)= S_k^z(t)$, with the same number of iterations.
Biquadratic interactions  can be treated by the same
iterative scheme, but inclusion of three-spin interaction 
would require reconsideration of the sublattice decomposition.

\section{Results and Comparisons}
For a quantitative analysis of the integration methods outlined
above we restrict ourselves to the Hamiltonian 
given by Eq.(\ref{H}) for $\lambda = 1$ in $d = 3$. The underlying
lattice is simple cubic with $L = 10$ lattice sites in each direction
and periodic boundary conditions in all cases discussed below.

In order to compare the different integration methods we first
investigate the accuracy within which the conservation laws are
fulfilled. The initial configuration is a well
equilibrated one from a Monte-Carlo simulation for $\lambda = 1$
at a temperature $T = 0.8T_c$ for $D=0$ and $D=J$, where $T_c$ refers
to the critical
temperature of the isotropic model $(D = 0)$. The magnetization of
such a configuration is non-zero and provides an indicator for
the numerical quality of the magnetization conservation. We integrate
the equations of motion to $t = 800/J$ and monitor the energy $e(t)
\equiv E(t)/(J L^3)$ of the configuration per spin and 
the modulus $m(t) \equiv
|{\bf M}(t)| / L^3$ of the magnetization per spin for 
the isotropic case $D = 0$ and its $z$-component $m_z(t) \equiv M_z(t)
/ L^3$ for the strongly anisotropic case $D = J$ as functions of time.
Note that for these tests both integration methods are started from
identical initial configurations. 

\begin{figure}[htbp]
\epsfxsize=3.0in
\begin{center}
\leavevmode
\epsffile{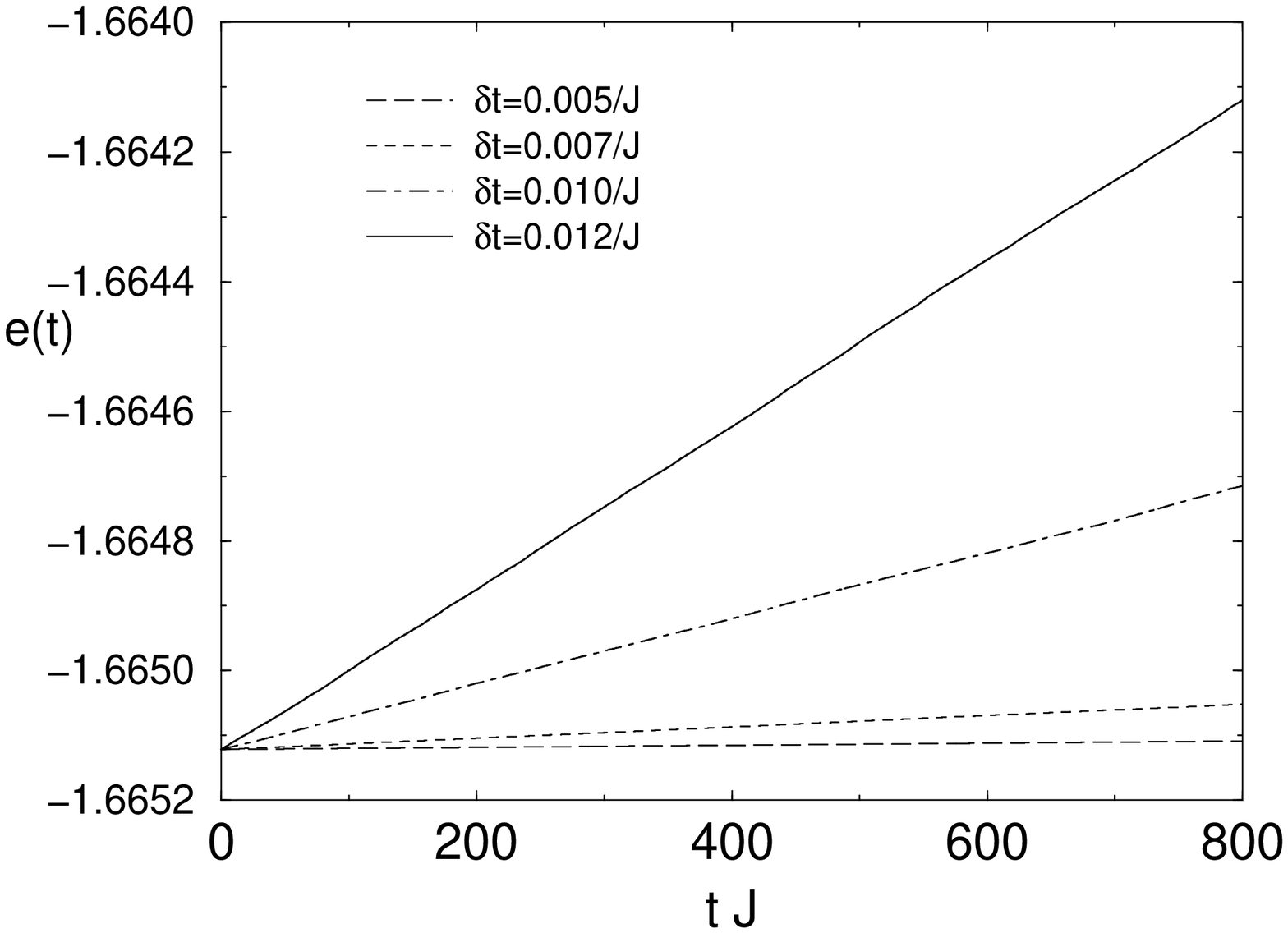}
\end{center}
%
\vspace{-1.0cm}
\fcaption{Energy $e(t) = E(t)/(J L^3)$ per spin for the predictor-corrector
method (see Eqs.(\protect\ref{AdBash}) and (\protect\ref{AdMoul})) for
several time steps and $D = 0$. 
\label{E2}}
\end{figure}

\begin{figure}[htbp]
\vspace{-2.5cm}
\epsfxsize=3.1in
\begin{center}
\leavevmode
\epsffile{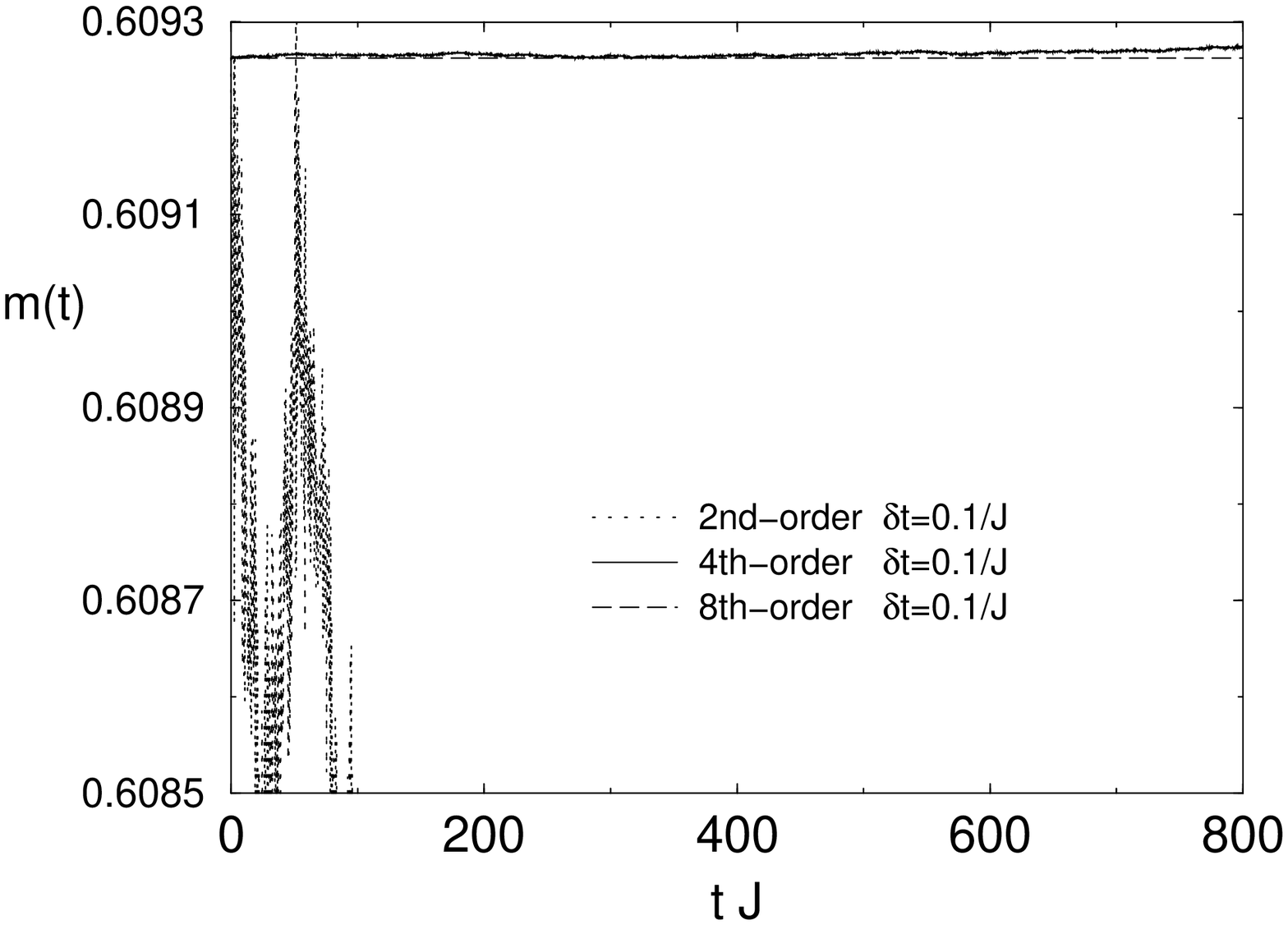}
\end{center}
\vspace{-1.0cm}
\fcaption{Magnetization $m(t) = |{\bf M}(t)|/L^3$ per spin for different order
decomposition schemes for $D=0$ and time step $\delta t = 0.1/J$: (dotted
line) 2nd-order scheme; (solid line) 4th-order scheme; (dashed
line) 8th-order method.
\label{Mz2}}
\end{figure}

\begin{figure}[htbp]
\vspace{-4.0cm}
\epsfxsize=3.1in
\begin{center}
\leavevmode
\epsffile{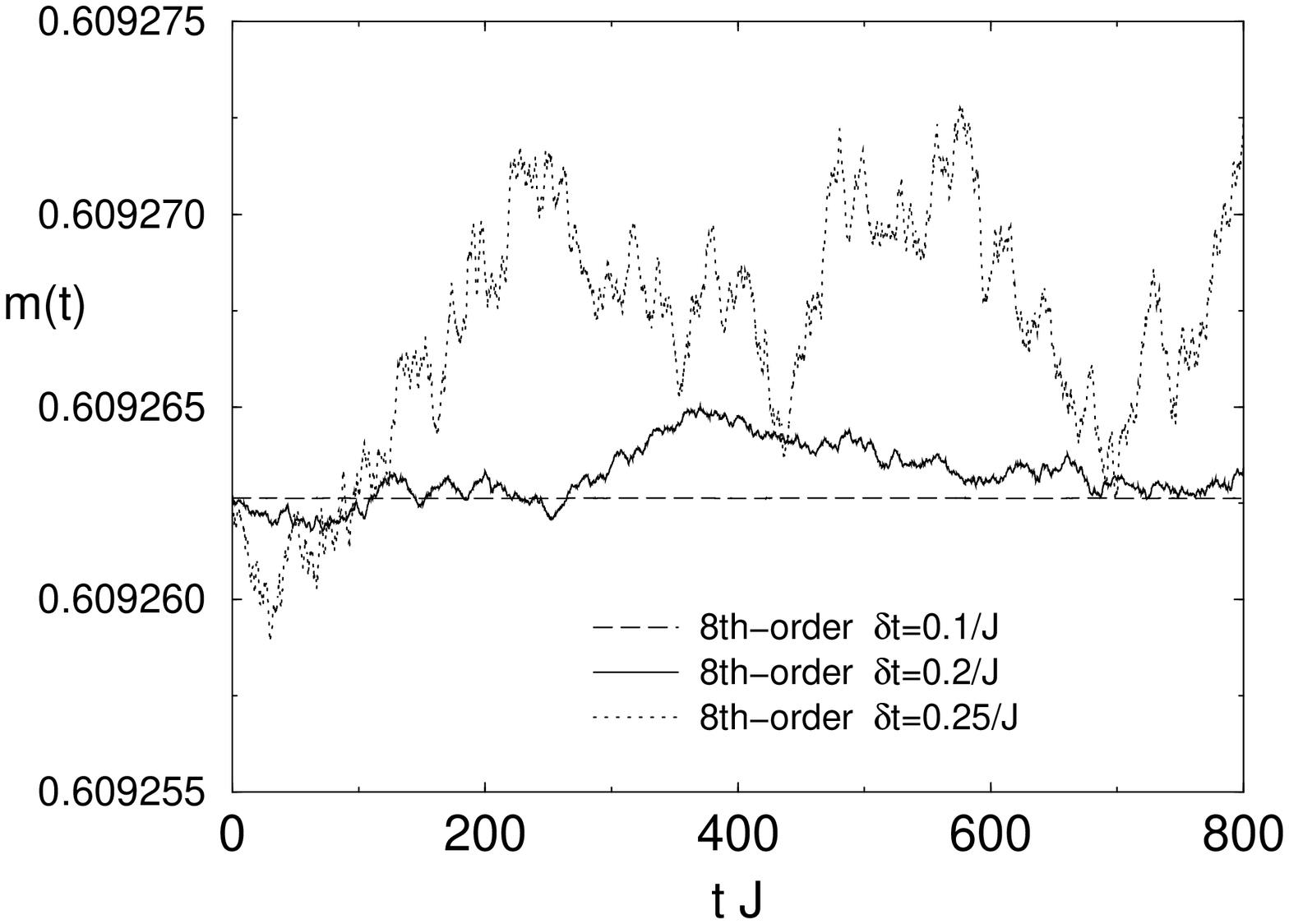}
\end{center}
\vspace{-1.0cm}
\fcaption{Magnetization $m(t)$ per spin for $D=0$ using the eighth-order decomposition
method (see Eq.(\protect\ref{epABA})) with different time steps.
\label{Mz8}}
\end{figure}

\begin{figure}[htbp]
\vspace{-2.0cm}
\epsfxsize=4.0in
\begin{center}
\leavevmode
\epsffile{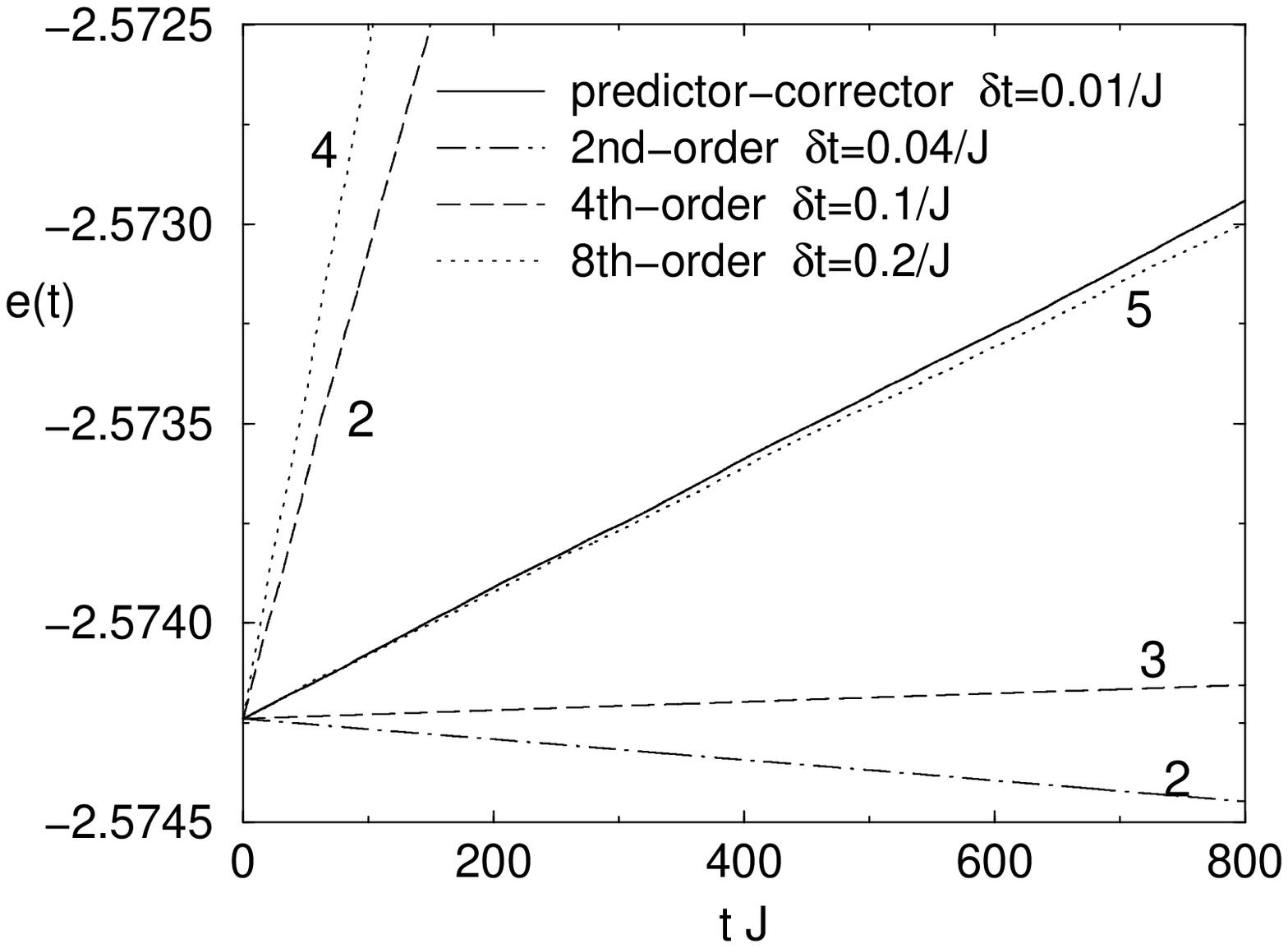}
\end{center}
%
\vspace{-1.5cm}
\fcaption{Energy $e(t)$ per spin for different order
decomposition schemes for $D=J$: (solid line) predictor-corrector method;
(dot-dashed line) 2nd-order scheme; (dashed line) 4th-order scheme; (dotted
line) 8th-order method. The number of iterations performed are marked next
to each line.
\label{e248J}}
\end{figure}

For $D = 0$ the implementation of
Eqs.(\ref{eA2BA2}) and (\ref{epABA}) using Eq.(\ref{Sktdt}) is
straightforward. The Taylor polynomial for $\sin x$ is chosen as
$\sin x = x - x^3/6$ for Eq.(\ref{eA2BA2}) and 
up to (and including) the terms $x^5/120$ and $x^9/9!$ for the
4th- and 8th-order decompositions in 
Eq.(\ref{epABA}), respectively, in order to reduce the 
magnetization fluctuations. Fig.\ref{E2} shows that $e(t)$ for
the predictor-corrector method increases linearly with time whereas 
the decomposition methods both yield $e(t) = const$.
Thus, $\delta t = 0.01/J$ is about the largest value that can be used without 
introducing substantial non-conservation of the energy.
Fig.\ref{Mz2} displays the magnetization conservation
for the 2nd-, 4th- and 8th-order decomposition methods, all with the same time
step $\delta t = 0.1/J$. The predictor-corrector method conserves $m(t)$
{\em exactly}, whereas the decomposition methods cause fluctuations of
$m(t)$ on all time scales.  It is also clear that the second-order method
is unstable with such a large time step.  The temporal structure of
$m(t)$ for the eighth-order decomposition methods with different time
steps is displayed in Fig.\ref{Mz8}. 
Even with a time step as large as
$\delta t = 0.25/J$, the magnetization is rather well conserved out to a
time of $t_{max} = 800 J^{-1}$.  
The three different decomposition methods
\vspace{-2.3cm}
%
%
%
can be made to produce fluctuations of the same magnitude, 
e.g. of $2\times 10^{-5}$,
by adjusting the
value of $\delta t$ to be $0.007 J^{-1}$ for the 2nd-order method, 
$0.1 J^{-1}$
for 4th-order and $0.25 J^{-1}$ for 8th-order.  
A single integration of the equations of
motion using Eq.(\ref{eA2BA2}) (2nd-order decomposition) is about
twice as fast as the predictor-corrector method. The
4th- and 8th-order decompositions (see Eq.(\ref{epABA})),
however, are respectively about $2.5$ and $9$ times slower than the 
predictor-corrector
method. Taking the change in time step by factors of $0.7$, $10$, and $25$,
respectively, into account, we find that the 2nd-, 4th- and 8th-order 
decomposition methods
yield a speedup of the integration of the equation of
motion by factors of approximately $1.5$, $4$, and $2.5$, respectively. 
Using time steps $\delta t=0.04/J$ and $0.2/J$ for the 2nd- and 4th-order
methods, respectively, yields in both cases an eightfold speedup as compared 
to the predictor-corrector method with $\delta t=0.01/J$.
If the overall quality of the magnetization 
conservation is also taken into account, there is a clear advantage for the 
4th-order decomposition according to Eq.(\ref{epABA}) for the isotropic
case $D = 0$. Somewhat surprisingly, with the current implementation,
the 8th-order method is not competitive.  For strong anisotropy, 
$D = J$, the
predictor-corrector method can be applied as before, but the
decomposition scheme must be modified because the spin
rotation axis depends on the spin value $S_k^z$ at the future time $t
+ \delta t$ (see Eq.(\ref{OmegaD})). 
As described in Sec. 2.2 this self consistency problem is solved
iteratively, where the quality of the energy conservation
depends on the number of iterations performed. For the 2nd-order
decomposition with $\delta t =
0.04/J$ two iterations are sufficient to obtain a better energy
conservation than the predictor-corrector method. Thus
a time integration step using the 2nd-order 
decomposition takes twice as long as for $D = 0$, so that its
advantage in speed only comes from the increase in the time step,
which is still a factor of four. For the 4th-order decomposition
with $\delta t = 0.2/J$ six iterations
are needed to obtain energy conservation to within six significant
digits, so one integration becomes 15 times slower than one with the
predictor-corrector method. From the increase of $\delta t$ by a
factor 20 only a 30\% gain in speed is obtained.
The number of iterations needed decreases with $\delta t$, but this
decrease does not compensate the loss in speed due to the smaller time
step. However, one still obtains a greatly improved energy
conservation.  All methods behave similarly, the
change in energy is basically linear with time (see Fig.\ref{e248J}). 
The reason for this is
the iterative nature of all four methods in the case $D \neq 0$. 
The overall
accuracy of the magnetization conservation appears to be independent
of $D$ for all decomposition methods. Considering both
overall energy conservation {\em and} speed, the 2nd-order
decomposition has some advantages over the
predictor-corrector method. 
%
%
%
%
%
\begin{figure}[htbp]
\epsfxsize=2.5in
\vspace{-2.0cm}
\begin{center}
\leavevmode
\epsffile{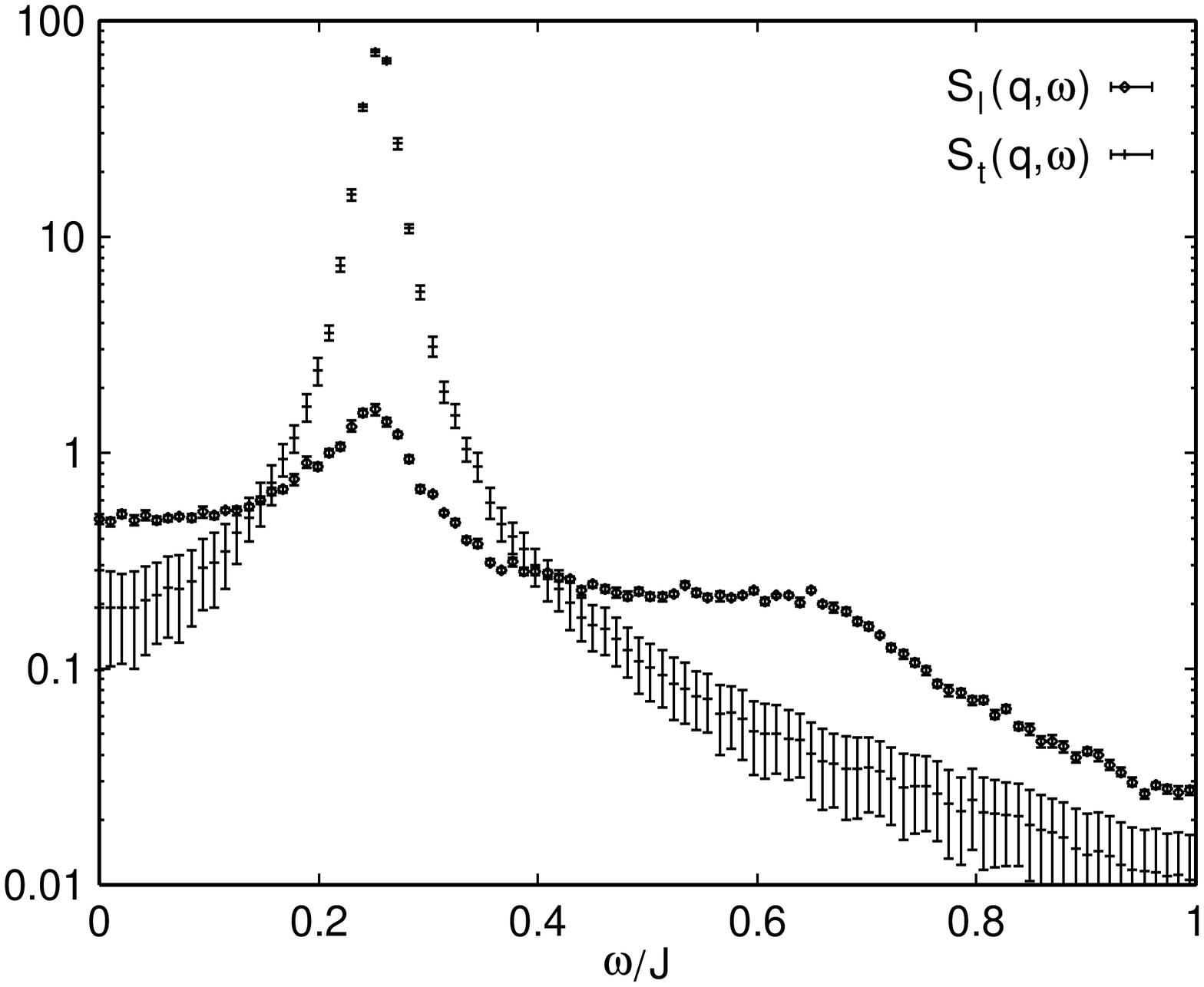}
\end{center}
%
\vspace{-1.0cm}
\fcaption{Dynamic structure factor of an
isotropic Heisenberg ferromagnet for $T = 0.8 T_c$ and $|{\bf q}| =
\pi/5$ in the $(100)$ direction on a simple cubic lattice ($L = 10$)
obtained with the 2nd-order decomposition method for time step
$\delta t = 0.04/J$. The longitudinal component is
$S_l({\bf q},\omega)$ and the transverse component is
$S_t({\bf q},\omega)$. \label{Sql}}
\end{figure}
%

Lastly, in Fig.\ref{Sql}
we show spin dynamics data for the dynamic structure factor 
in the $(100)$ direction for the isotropic Heisenberg
ferromagnet ($D = 0$) on a simple cubic lattice ($L = 10$) at $T = 0.8
T_c$ obtained with the second-order decomposition method (see
Eq.(\ref{eA2BA2}), $\delta t = 0.04/J$).  The equations
of motion have been integrated to $800/J$ and averages have been taken
over 1000 initial configurations, where the time displaced correlation
functions have been measured to $400/J$.  A spin wave peak
is located at $\omega_0 = 0.25 J$ and the
shoulder like feature at $\omega \simeq 0.5 J$ in $S_l({\bf
q},\omega)$ (see Fig.\ref{Sql}) is due 
to multi-spin-wave processes, the
description of which is beyond the scope of this article. 

\section{Summary}
We have described a set of algorithms which is based on
Suzuki-Trotter decompositions of exponential operators and compared their 
relative performance with each other as well as with a  
predictor-corrector method.  The
advantages of the predictor-corrector method are its
versatility and its capability to conserve the magnetization exactly.  
The decomposition of the lattice into sublattices, which is
the basis for the decomposition method, depends on the range of the
interactions, so that this approach is less general than
the predictor-corrector method. Crystal field anisotropies leave the
performance of the predictor-corrector method almost unaffected,
whereas the decomposition method suffers from a drastic reduction in
speed.

The advantage of the decomposition method is its ability to
handle large time steps and to conserve spin
length exactly. In the absence of anisotropies it also conserves
the energy exactly and it maintains reversibility. For anisotropic
Hamiltonians energy conservation and reversibility can be obtained to
a high accuracy using iterative schemes. Exact magnetization
conservation, however, is lost. The time steps which can be used far
exceed those used by the predictor-corrector method.  In simple cases 
the 4th-order
decomposition yields very accurate results even for time steps,
which are an order of magnitude larger than typical time steps used
for the predictor-corrector method.  The 8th-order algorithm
improves the conservation significantly but at the cost of greatly 
increased execution time.

\nonumsection{Acknowledgments}
\noindent
M. Krech gratefully acknowledges financial support of this work
through the Heisenberg program of the Deutsche Forschungsgemeinschaft.
This research was supported in part by NSF grant \#DMR - 9727714.

\nonumsection{References}

\end{document}